\documentclass[english,aps, twocolumn]{revtex4}
\usepackage[T1]{fontenc}
\usepackage[latin9]{inputenc}
\usepackage{color}
\usepackage{amsmath}
\usepackage{graphicx}
\usepackage{esint}

\makeatletter
\@ifundefined{textcolor}{}
{%
 \definecolor{BLACK}{gray}{0}
 \definecolor{WHITE}{gray}{1}
 \definecolor{RED}{rgb}{1,0,0}
 \definecolor{GREEN}{rgb}{0,1,0}
 \definecolor{BLUE}{rgb}{0,0,1}
 \definecolor{CYAN}{cmyk}{1,0,0,0}
 \definecolor{MAGENTA}{cmyk}{0,1,0,0}
 \definecolor{YELLOW}{cmyk}{0,0,1,0}
}

\makeatother

\usepackage{babel}
\begin{document}

\title{Quantum Transport in Graphene Nanoribbons with Realistic Edges}

\author{Patrick Hawkins, Milan Begliarbekov, Marko Zivkovic, Stefan Strauf,
Christopher P. Search}

\email{mbegliar@stevens.edu}

\selectlanguage{english}%

\affiliation{Department of Physics and Engineering Physics, Stevens Institute
of Technology, Hoboken NJ 07030, USA}
\begin{abstract}
Due to their unique electrical properties, graphene nanoribbons (GNRs)
show great promise as the building blocks of novel electronic devices.
However, these properties are strongly dependent on the geometry of
the edges of the graphene devices. Thus far only zigzag and armchair
edges have been extensively studied. However, several other self passivating
edge reconstructions are possible, and were experimentally observed.
Here we utilize the Nonequilibrium Green\textquoteright{}s Function
(NEGF) technique in conjunction with tight binding methods to model
quantum transport through armchair, zigzag, and several other self-passivated
edge reconstructions. In addition we consider the experimentally relevant
cases of mixed edges, where random combinations of possible terminations
exist on a given GNR boundary. We find that transport through GNR's
with self-passivating edge reconstructions is governed by the sublattice
structure of the edges, in a manner similar to their parent zigzag
or armchair configurations. Furthermore, we find that the reconstructed
armchair GNR's have a larger band gap energy than pristine armchair
edges and are more robust against edge disorder. These results offer
novel insights into the transport in GNRs with realistic edges and
are thus of paramount importance in the development of GNR based devices.
\end{abstract}

\keywords{Keywords: Edge Reconstruction, Graphene Nanoribbon, Quantum Transport,
Edge Disorder, Green's Functions, Band Gap}

\maketitle

\section{Introduction}

The study of quantum transport in graphene nanoribbons (GNR's) is
of particular interest for the development of novel nanoelectronic
devices since, unlike in conventional materials, the transport characteristics
of nanostructured graphene based devices are affected by the geometry
(chirality) of the GNR edges. In the ideal case the edges of GNR's
can be arranged in two distinct geometries: zigzag and armchair, as
shown in Fig. 1A. Pristine armchair edges, i.e., armchair edges with
no atomic roughness, are semiconducting \cite{Malard2009,Nakada1996},
while pristine zigzag edges are metallic \cite{Malard2009,Nakada1996},
as shown in Figs. 1B and 1C., in which the current density through
zigzag and armchair terminated GNR's is plotted, as calculated using
the methodology described below. The homogenous current density in
the zigzag GNR is a signature of the metallic density of states at
the edges. Conversely, the current density of the armchair terminated
GNR is inhomogeneous and shows several distinct paths of electronic
carriers, known as quantum billiards \cite{Ponomarenko2008,Miao2007},
which result from the reflection of carriers off of the semiconducting
edges. The semiconducting nature of the armchair edge originates from
a quantum confinement effect, whereas the metallic nature of the zigzag
edge stems from a nondispersive state localized at the periphery of
the GNR \cite{Wimmer2010,Begliarbekov2011a}. The localized states
stems from the fact that the bottom of the conduction band and the
top of the valence band are always degenerate at the $k=\pi$ point
in the Brilluoin zone. The origins of this degeneracy are discussed
in detail in Ref. \cite{Nakada1996}. 

\begin{figure}
\includegraphics[scale=0.21]{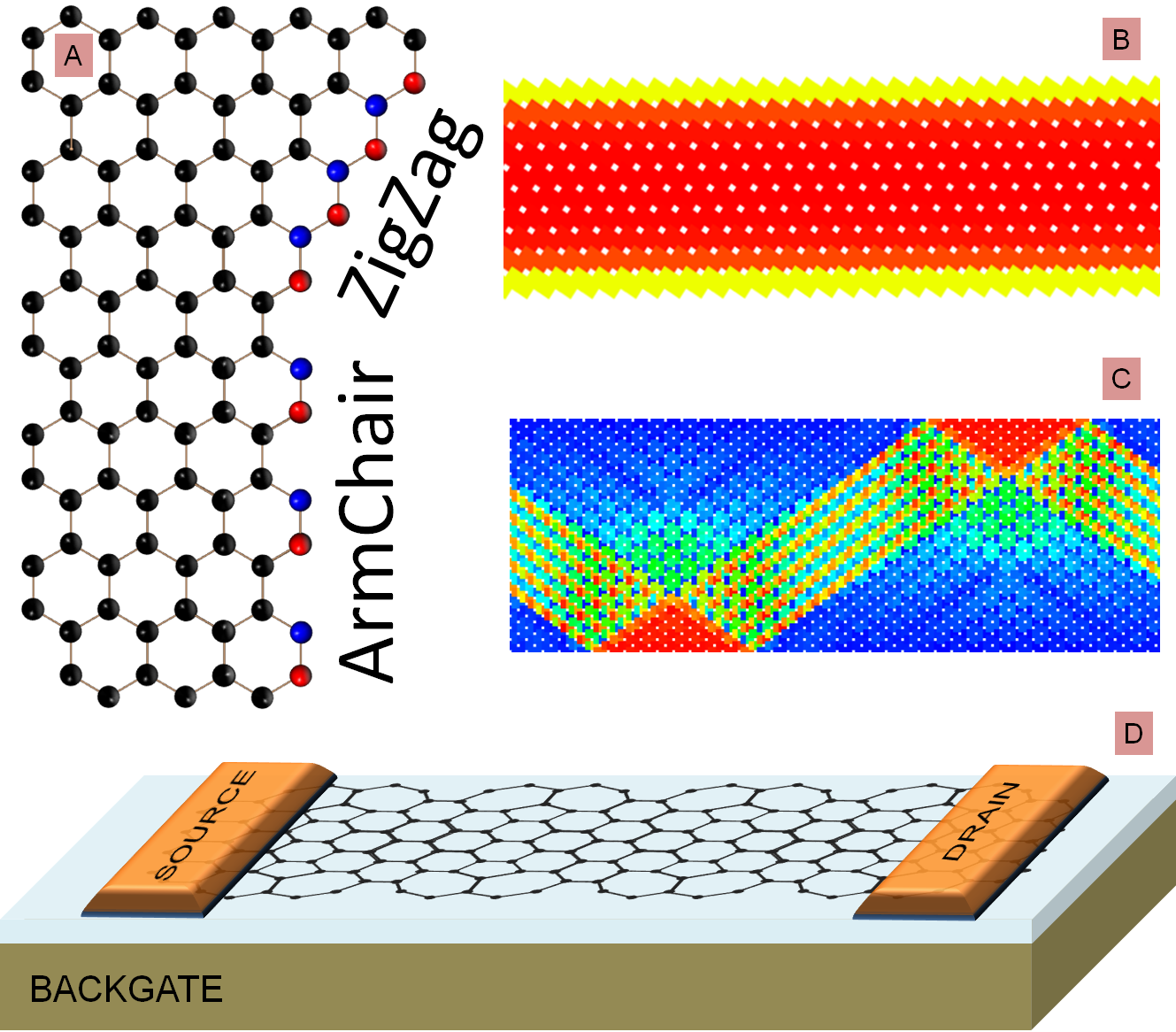}

\caption{\textbf{A} A schematic of a graphene lattice highlighting the ideal
armchair and zigzag edges. The edge atoms are colored red and blue
to reflect the valleys to which they belong. Contour plots of the
2D current density through perfect \textbf{B} zigzag and \textbf{C}
armchair nanoribbons, for the device geometry shown in panel \textbf{D}.
The reddish (bluish) colors correspond to high (low) current densities.
Both ribbons are 10 nm x 30 nm, backgate $V_{bg}$ bias is 0.01 eV,
and an infinitesimal source drain bias.}
\end{figure}

The contrasting behavior of chiral zigzag and armchair edges allows
the possibility of utilizing armchair and zigzag GNR's for different
electronic and photonic applications. For example, armchair GNR's
are ideal for use in high speed transistor \cite{Lin2010,Begliarbekov2011b}
and photodetector \cite{Xia2009} devices that require a bandgap.
The localized state of the zigzag GNR's is ferromagnetic \cite{Jiang2007}
and is thus capable of rendering the GNR half-metallic under the application
of a lateral electric field \cite{Son2006}. Consequently, zigzag
GNR's are ideal for spintronic applications. Furthermore combining
zigzag and armchair edges in a single device results in novel behaviors
such as current rectification in Z-shaped GNR's \cite{Wang2007,Wang2008}
and spin filtering \cite{Saffarzadeh2011}.

Although extensive theoretical models have been put forth to study
transport through pristine armchair and zigzag GNR's, \cite{Wakabayashi1999}
little is known about transport through GNR's with realistic edges,
i.e., edges that are not atomically pristine. Since all experimentally
realized GNR's have thus far possessed some degree of edge roughness
\cite{Begliarbekov2010,Begliarbekov2011}, an understanding of transport
through these devices is vital for the design of novel electronic
devices. Furthermore, while ideal armchair and zigzag edges are the
most frequently studied edge types, other self-passivating edge reconstructions
are possible \cite{Koskinen2008} and have been experimentally observed
\cite{Koskinen2009,Girit2009}. The influence of partially and fully
reconstructed edges on transport in GNR's has not yet been analyzed
in detail. In this work we utilize nonequilibrium Green's functions
(NEGF) methods to numerically simulate quantum transport through GNR's
with both atomically smooth as well as rough edges. We furthermore
simulate transport in self-passivated GNR's with various degrees of
edge roughness. We find that transport through GNR's with self-passivating
edge reconstructions is governed by the sublattice structure of the
edges, in a manner similar to their parent zigzag or armchair configurations.
Furthermore, we find that the reconstructed armchair GNR's have a
larger band gap energy than pristine armchair edges and are more robust
against edge disorder. Taken together our results elucidate the effects
of edge reconstructions on the transport properties of GNR's and are
thus of paramount importance for the design of novel electronic and
photonic devices.

\subsection{Model}

Numerical simulations have been carried out with the aid of nonequilibrium
Green's functions (NEGF) in conjunction with the tight binding formalism,
where only the first order (nearest neighbor) hopping is considered.
The simulations were implemented using the KNIT algorithm \cite{Kazymyrenko2008}
on a 46 node parallel cluster. The system Hamiltonian is of the form 

\begin{equation}
\hat{H}=\underset{i\neq j}{\sum}\gamma_{ij}c_{i}^{\dagger}c_{j}+\underset{i=1}{\sum}\epsilon_{i}c_{i}^{\dagger}c_{j},
\end{equation}

\noindent where $\gamma_{ij}$ is the nearest neighbor hopping integral,
$\epsilon_{i}$ is the on-site energy, and $c_{i}^{\dagger}$ / $c_{j}$
are the creation / annihilation operators for the $i$-th and $j$-th
lattice sites respectively. With the aid of the above Hamiltonian
the local current $I_{ij}$ and density of states $\rho_{i}$ can
be computed using the standard NEGF formalism according to

\[
I_{ij}=\int dE\left[\gamma_{ij}G_{ji}^{<}\left(E\right)-\gamma_{ij}G_{ij}^{<}\left(E\right)\right]\textrm{ and}
\]

\begin{equation}
\rho_{i}=\frac{1}{2\pi}\Im\left\{ \int dEG_{ii}^{<}\left(E\right)\right\} ,
\end{equation}

\noindent where $G_{ij}^{<}\left(E\right)$ is the retarded nonequilibrium
Green's function given by 

\[
G_{ij}^{<}\left(E\right)=i\int dte^{-iEt}\left\langle c_{j}^{\dagger}c_{i}\left(t\right)\right\rangle .
\]

The systems considered in this study consist of narrow graphene nanoribbons,
typically $\sim$10 x 30 nm connected to source-drain electrodes via
Ohmic contacts. The Fermi energy $E_{F}$ may be adjusted via the
application of a backgate bias \cite{Miao2007} as shown in Fig. 1d.
The above formalism is sufficient to describe transport through ideal
armchair and zigzag GNR's. However, in order to properly account for
the altered bond lengths of the self-passivated edge reconstructions
\cite{Koskinen2008}, the environmentally dependent tight binding
model was utilized \cite{Tang1996}, in which the change in the bond
lengths is accounted for by modifying the values of the hopping integrals
$\gamma_{i}$ according to

\begin{equation}
\gamma_{i}\left(R_{jk}\right)=\alpha_{1}R_{jk}^{-\alpha_{2}}\exp\left(-\alpha_{3}R_{jk}^{-\alpha_{4}}\right)\left[1-S_{jk}\right],
\end{equation}

\noindent where $\alpha_{i}$ are the scaling parameters and $S_{jk}$
are the screening functions calculated from first principles calculations
in Ref. \cite{Tang1996}. $R_{jk}$ are the bond lengths of the edge
reconstructions, which are calculated in Ref. \cite{Koskinen2008}.
Following the results of Ref. \cite{Tang1996}, the screening function
is of the form

\[
S_{kj}=\frac{\exp\left(\xi_{jk}\right)-\exp\left(-\xi_{jk}\right)}{\exp\left(\xi_{jk}\right)+\exp\left(-\xi_{jk}\right)},
\]

\noindent with

\begin{equation}
\xi_{jk}=\beta_{1}\underset{l}{\sum}\exp\left[-\beta_{2}\left(\frac{R_{jl}+R_{kl}}{R_{jl}}\right)^{\beta_{3}}\right],
\end{equation}

\noindent where $\beta_{i}$ are the screening, and $R_{jl}$ and
$R_{kl}$ are the next-nearest neighbor bond lengths. The bond lengths
and the values of the nearest neighbor hopping integrals are listed
in Fig. 2f.\textcolor{red}{{} }

\begin{figure*}
\includegraphics[scale=0.39]{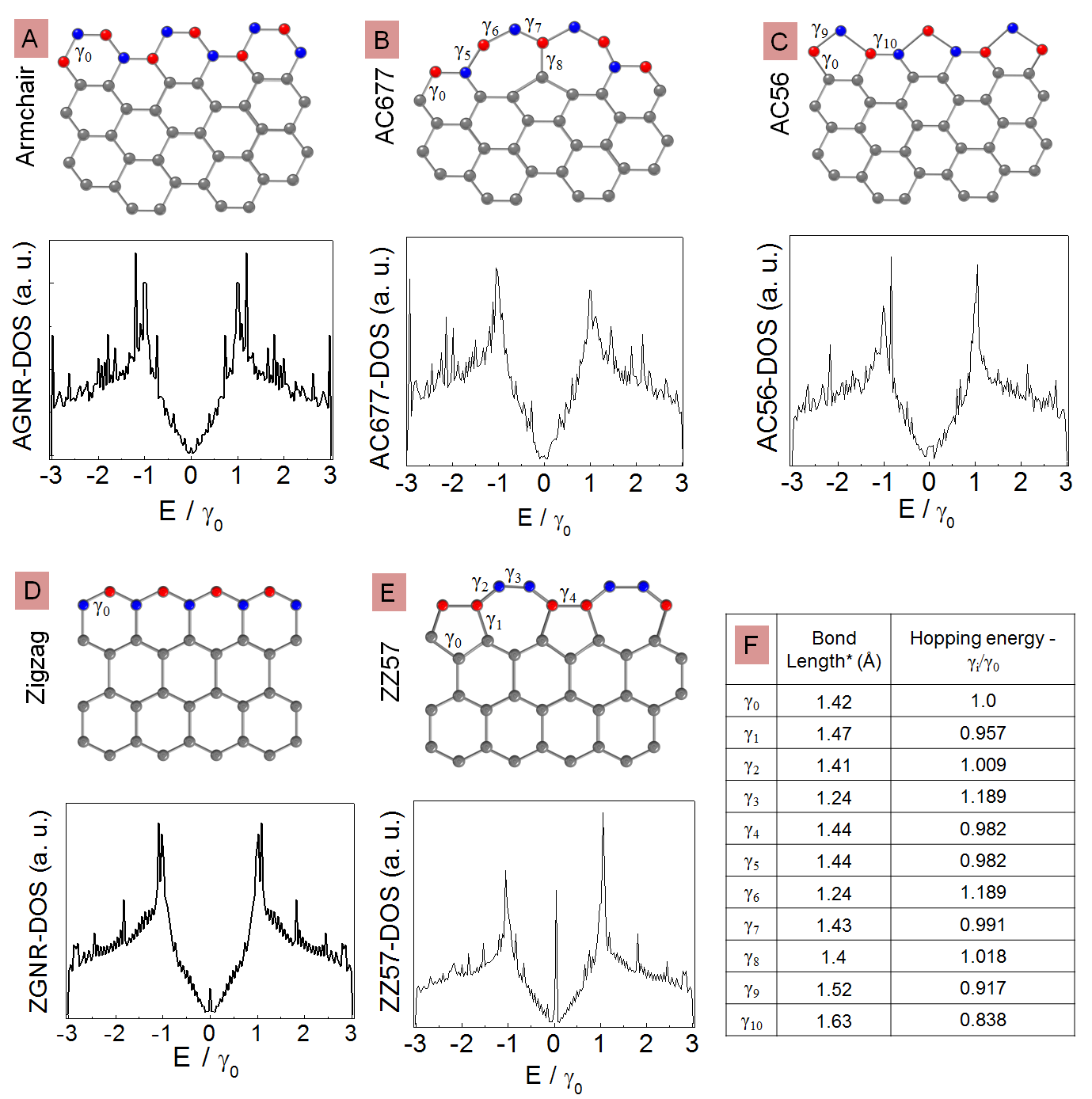}

\caption{Illustrations of the edge reconstructions and their densities of states
relative to the Fermi energy for the \textbf{A} armchair, \textbf{B}
AC677, \textbf{C} AC56 \textbf{D} zigzag, and \textbf{E} ZZ57 nanoribbons.
The table in \textbf{F} shows the values of the bond lengths (adopted
from Ref. 22) and change in the hopping energy $\gamma_{i}/\gamma_{0}$,
where the values of $\gamma_{i}$ $\left(i=1,2\cdots10\right)$ are
indicated in the diagrams in panels a-e. In these simulations we used
$\gamma_{0}=2.8$ eV.}
\end{figure*}

Before proceeding to the discussion of our results, it should be noted
that spin effects were previously predicted to play an important role
\cite{Yazuev2010,Girao2011} in transport of pristine zigzag GNRs
whose edge states result in flat bands at the Fermi energy. Spin effects
may be included by adding a Hubbard term $H'=U\underset{i}{\sum}n_{i\uparrow}n_{i\downarrow}$,
to the Hamiltonian where $U$ is a parameter that defines the strength
of the on-site Coulomb interaction. Previous calculations with zigzag
GNRs have shown that $H'$leads to the development of ferromagnetic
spin orientations along each edge and antiferromagnetic correlations
between the two edges. The spin polarization of the edge states breaks
the degeneracy of the flat energy bands at the Fermi energy with a
gap opening up at $E_{F}$. This band gap leads to a vanishing of
the density of states at $E_{F}$ while higher energy states are unaffected
by $H'$\cite{Yazuev2010,Song2010,Kunstmann2011}. There is some experimental
evidence for the energy splitting of the edge states \cite{Pan2012,Rao2012,Tao2011}although
there is a debate as to whether such magnetism could exist in real
world conditions \cite{Kunstmann2011}. Among the many effects that
would quench this edge state magnetism is edge reconstruction. Edge
reconstructed zigzag GNRs are expected to show no magnetic effects
since the edge state energy bands at $E_{F}$ are no longer flat \cite{Kunstmann2011,Song2010}.
Here we do not consider nonlinear spin effects in our results.

\section{Results and Discussion}

Figure 2 shows the atomic arrangements of the five types of edges
considered in this work: the armchair and zigzag edges as well as
the three types of self-passivating reconstructions, and their corresponding
densities of states. Although pristine armchair and zigzag edges are
expected for the honeycomb lattice geometry of graphene, thermal \cite{Xu2011}
and optical \cite{Begliarbekov2011} edge reconstruction has been
observed to give rise to the self passivating variants of the pristine
edges. The most common mechanism of bond rearrangement in sp2 crystalline
carbons is the Stone-Wales mechanism, which in the case of GNR edges
can lead to the ZZ57 reconstruction. This mechanism has been studied
extensively in relation to nanocrystalline carbon systems \cite{Ertekin2009,Ma2009,Fan2010}.
Furthermore, the energetics of other self-passivating edge reconstructions
were also investigated \cite{Koskinen2008}. The names of the reconstructed
edges originate from the parent edge type, i.e., zigzag or armchair,
and the geometry of the closed structure. For example, the ZZ57 edge
is a zigzag edge that is terminated by alternating pentagons and heptagons.
As can be seen in Fig. 2, the self passivating edge reconstructions
together with the unaltered edges form two distinct families of edges:
the zigzag and the armchair family. Although the reconstructed edges
bear little resemblance to their parent edge types, the zigzag reconstruction
is similar to the pristine zigzag edge in that it possesses a metallic
state at $E=0$, as can be seen in Figs. 2D and 2E. Interestingly,
the magnitude of the $E=0$ state is found to be larger for the ZZ57
reconstruction than for the zigzag GNR of the same width. Consequently,
ZZ57 edges might be of greater utility for devises that exploit the
edge localized states, such as devices based on graphene antidot superlattices
\cite{Begliarbekov2011a}. Phenomenologically, the increase in the
magnitude of the $E=0$ state can be attributed to Anderson localization
that arises from the additional break in the periodicity of the hexagonal
lattice at the ZZ57 edge, as can be seen in Figs. 2D and 2E. Similarly
both of the armchair reconstructions are semiconducting and possess
a finite energy gap as discussed in detail below. The similarities
between the parent edge types and their reconstructed counterparts
can be readily understood by considering the sublattice structure
of the terminating atoms at the edges. Both zigzag and ZZ57 edges
are terminated by atoms that belong to the same sublattice, whereas
armchair as well as AC677 and AC56 edges contain atoms that belong
to both sublattices. The terminating atoms are atoms that are the
outermost atoms of the GNR. Consequently, the armchair edges support
intervalley scattering, whereas the zigzag edges do not. Thus the
sublattice structure of the terminating atoms in the self passivating
edge reconstructions determines the properties of the resultant edges
in a manner identical to the pristine armchair and zigzag edges.

\begin{figure}
\includegraphics[scale=0.22]{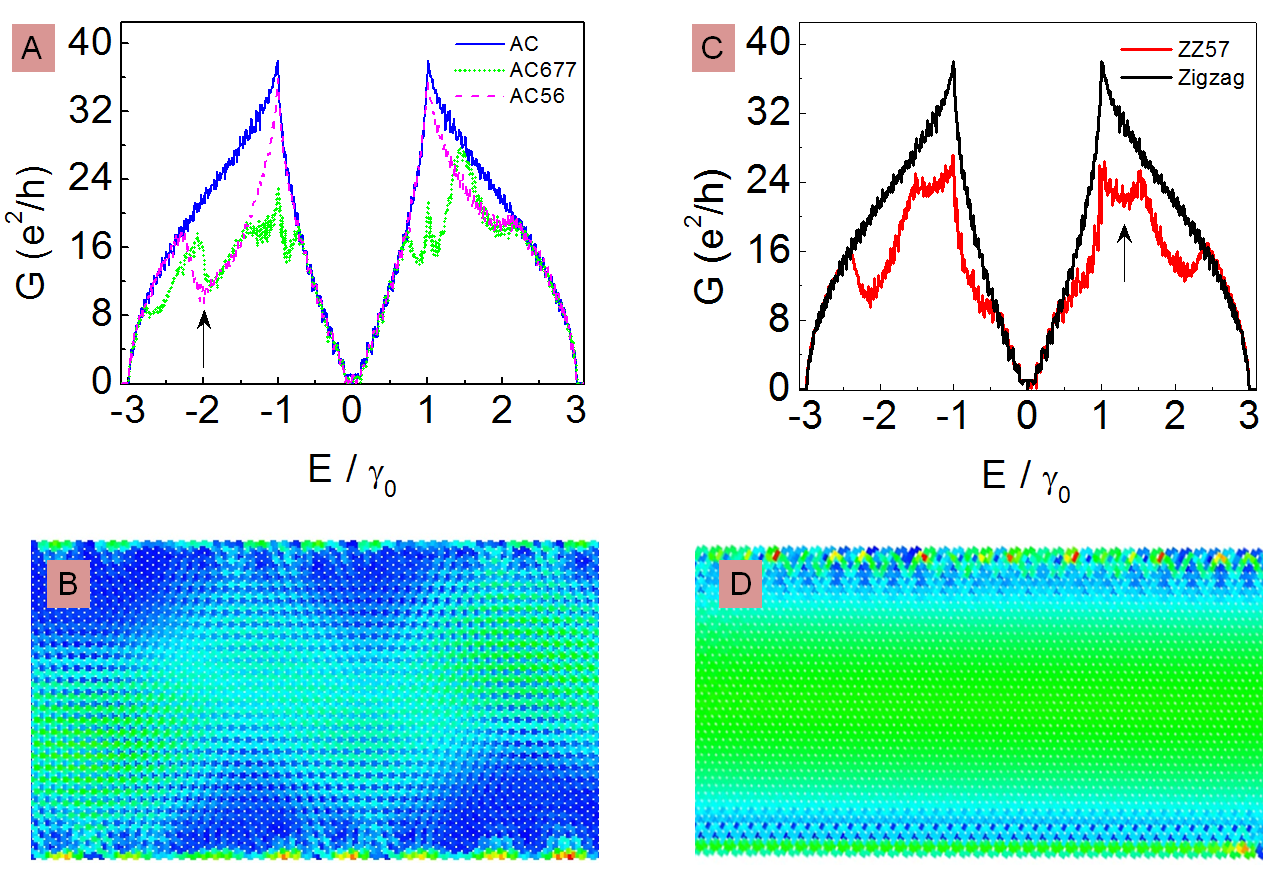}

\caption{\textbf{A} Conductance of ideal armchair (solid blue), AC677 (dash-dotted
green), and AC56 (dashed pink) lines. \textbf{C} Conductance through
ideal zigzag (solid black) and ZZ57 (dotted red) lines. The 2D current
density for \textbf{B} AC56 and \textbf{D} ZZ57 reconstructions for
the same parameters as in Fig. 1.}

\end{figure}

Having discussed the principle features of the zigzag and armchair
families we now turn to the discussion of transport through GNR\textquoteright{}s
with self passivating edge reconstructions. We first restrict the
discussion to fully reconstructed edges. Figures 3A and 3C show the
conductance of pristine and reconstructed armchair and zigzag terminated
GNR's respectively. Figures 3B and 3D show the 2-D current density
of AC56 and ZZ57 edges. Furthermore the 2-D current density for the
AC677 reconstruction was found to be similar to that of the AC56 edge
(data not shown). The prominent features in the conductance through
GNRs with reconstructed edges are the dips in the conductance at certain
values of the Fermi energy. These dips arise from the fact that the
size of the unit cell is enlarged at the reconstructed edges. Consequently,
fewer states are available per unit area as compared to the unreconstructed
edges resulting in pronounced dips in the conductance. It should be
further noted that these dips appear energetically far away from the
Dirac point. These features can potentially be probed in either ARPES
or optical conductivity experiments. However, since no dips are observed
in the vicinity of the Dirac point, the low bias conductance of reconstructed
GNR's should be similar to the conductance of pristine GNR's. Therefore,
our results demonstrate that while transport in narrow GNR's is highly
sensitive to the type of edge terminations, namely, zigzag or armchair,
it is robust with regard to self-passivating edge reconstructions
within the same edge family. Consequently, the relevant physical mechanisms
that dominate transport in armchair and zigzag terminated GNR's can
be still observed in GNR's with realistic edge reconstructions. 

The 2D current density of AC56 and ZZ57 edges further highlights the
similarity between the reconstructed and pristine GNR edges. The AC56
terminated GNR's show an oscillatory current density characteristic
of quantum billiards in pristine armchair GNRs, whereas the current
density in the ZZ57 terminated GNR is homogenous. The most striking
difference in the 2D current density between pristine and reconstructed
edges is the presence of edge-localized charge density. Phenomenologically
this effect can be understood in terms of Anderson localization since
the reconstructed edges break the periodicity of the underlying lattice
thereby localizing the electronic wavefunctions. As a result the quantum
billiard patterns are smeared by the edge-localized modes. 

\begin{figure}
\includegraphics[scale=0.27]{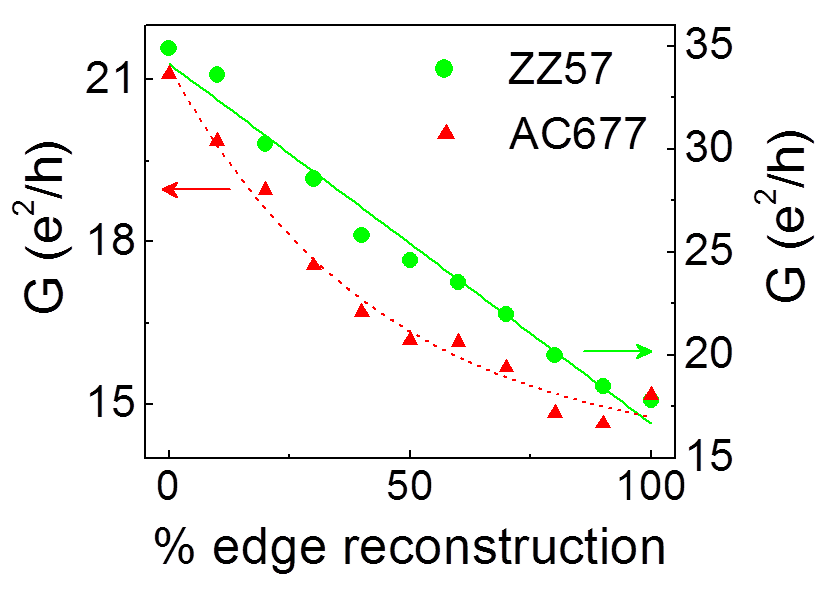}

\caption{Conductance of an armchair (red triangles) and a zigzag (green circles)
nanoribbons with varying percentages of edge reconstruction with AC677
and ZZ57 substitutions respectively. The conductance values were extracted
for $E_{F}=2.1$ and $E_{F}=1.1$ eV for the armchair and zigzag nanoribbons
respectively, as indicated by the arrows in Fig. 3a and 3c, for the
same size devices as indicated in Fig. 1. }
\end{figure}

Having analyzed the effects of complete reconstruction on transport
through GNR's we now turn our attention to the case of partial reconstructions.
To study the effects of partial reconstructions on transport through
GNR's we first calculate the conductance of a pristine armchair and
zigzag GNR and then introduce various percentages of AC677 and ZZ57
edges respectively. To quantify the effect of reconstruction, we plot
the conductance value of two dips ($E=2.1$ eV for the armchair reconstruction
and $E=1.1$ eV for the zigzag reconstruction) as shown in Fig. 4.
The dips are identified by black arrows in Figs. 3Aand 3C. As can
be seen in Fig. 4 the onset of the dips occurs at small percentages
of edge reconstructions ($\sim$5\%). Increasing the amount of edge
reconstruction increases the amplitude of the dips, which become most
pronounced for fully reconstructed GNR's. It should be noted that
the armchair reconstructions have a more dramatic effect on the conductivity
than the zigzag reconstructions. The magnitude of the dip in the conductance
of an armchair GNR reconstructed by AC677 segments decreases exponentially
according to $G\left(\%x_{\textrm{AC667}}\right)=A_{1}\exp\left(-\%x_{\textrm{AC667}}/A_{2}\right)+A_{3}$,
where $\%x_{\textrm{AC667}}$ is the percentage of AC677 segments
and $A_{1}=8.1$, $A_{2}=44.1$, and $A_{3}=13.0$ are empirical fit
parameters in units of $e^{2}/heV$. The magnitude of the dip in the
conductance of a zigzag GNR decreases linearly as a function of ZZ57
reconstruction according to $G\left(\%x_{\textrm{ZZ57}}\right)=B_{1}\left(\%x_{\textrm{ZZ57}}\right)+B_{2}$,
where $\%x_{\textrm{ZZ57}}$ is the percentage of ZZ57 reconstruction
and $B_{1}=-0.17$ and $B_{2}=34.2$ are empirical fit parameters
in units of $e^{2}/heV$.

\begin{figure}
\includegraphics[scale=0.37]{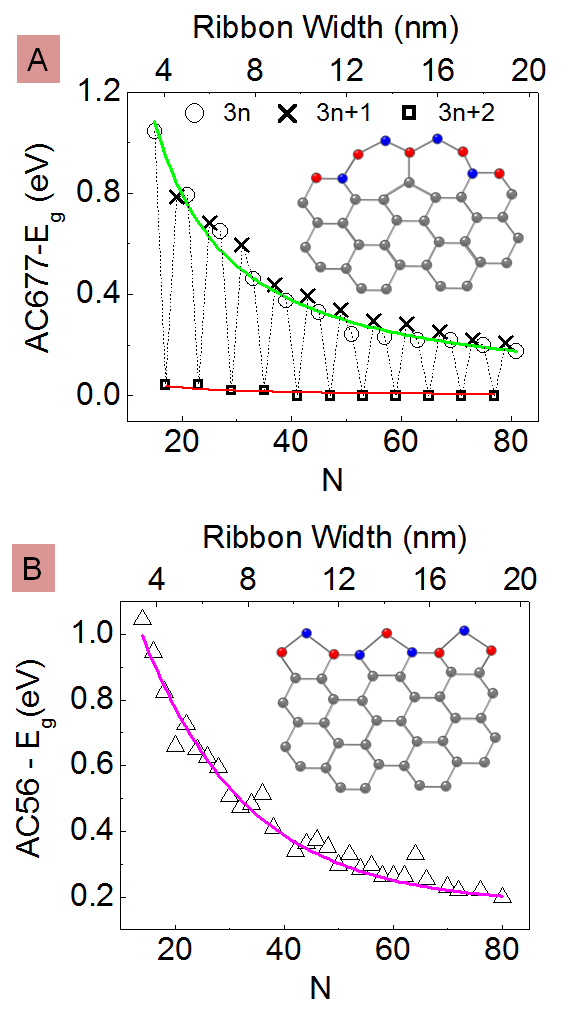}

\caption{Bandgap as a function of ribbon width for \textbf{A} AC677 and \textbf{B}
AC56 nanoribbons, for the same device parameters as indicated in Fig.
1. $n$ is the number of Carbon atoms that span the width $N$ of
the GNR. The different symbols in panel 'a' represent the multiplicity
of the number of atoms that span the width of the GNR. The circles,
crosses, and squares indicate GNR's of widths that are multiples of
$N=3n$, $N=3n+1$, and $N=3n+2$ respectively.}
\end{figure}

Finally, we turn our attention to the bandgap of reconstructed armchair
GNR's. In light of the above discussion it is sufficient to only consider
fully reconstructed GNR's since only the absolute magnitude of the
bandgap is effected by partially reconstructed GNR's. The gap energy,
$E_{G}$, is extracted from the conductance of the GNR. Figures 5a
and 5b show the magnitude of $E_{G}$ as a function of ribbon width
for AC677 and AC56 ribbons respectively. As can be seen in Fig. 4,
the bandgap of very narrow GNR's less than 4 nm surpasses 1 eV; however,
in both cases it decreases exponentially as a function of ribbon width.
For the AC677 terminated ribbon the gap energy as a function of ribbon
width $E_{G}^{(AC667)}\left(n\right)$ decreases according to $E_{G}^{(AC667)}(n)=C_{1}\exp\left(-n/C_{2}\right)+C_{3}$,
where $n$ is the number of atoms that span the width $N$ of the
GNR, and $C_{1}=2.1$, $C_{2}=17.5$, and $C_{3}=0.2$ are empirical
fit parameters in units of $e^{2}/heV$. Similarly, $E_{G}^{(AC56)}\left(n\right)$
for the AC56 ribbons varies according to $E_{G}^{(AC56)}(n)=D_{1}\exp\left(-n/D_{2}\right)+D_{3}$,
where $D_{1}=1.7$, $D_{2}=17.8$, and $D_{3}=0.2$ are empirical
fit parameters in units of $e^{2}/heV$. A striking difference between
the AC677 and AC56 GNR's is that the AC677 possess an additional family
dependence of the gap energy on the multiplicity of the number of
atoms that span the width GNR. Namely, AC677 terminated GNR's with
width's that are multiples of $3n$ and $3n+1$ possess an electronic
bandgap, whereas AC677 terminated GNR's with widths that are multiples
of $3n+2$ do not have a bandgap. The same family dependence on the
width multiplicity was previously calculated for pristine armchair
terminated GNR's \cite{Yang2007,Wakabayashi2010,Zheng2007}. This
behavior, however, vanishes in AC56 terminated GNR's. The lack of
the multiplicity dependence of $E_{G}$ in AC56 terminated GNR's can
be understood from the fact that the AC56 geometry imposes a greater
strain on the underlying lattice, thereby distorting more of the features
of pristine armchair GNR's. Finally, it is interesting to note that
$E_{G}$ in both AC677 and AC56 terminated ribbons is twice as large
as $E_{G}$ in pristine armchair nanoribbons of the same width (see,
for example, the numerical calculations of the armchair bandgap in
Ref. \cite{Yang2007}). Consequently, these variants of the armchair
GNR's might become more useful candidates for GNR based devices, if
a reliable method for fabricating these edges is found.

\section{Summary}

In summary, we simulated quantum transport through graphene nanoribbons
with realistic edge terminations by utilizing nonequilibrium Green's
functions. We found that transport in the self-passivated edge reconstructions
is qualitatively similar to transport in the parent edge type. This
similarity is attributed to the sublattice structure of the terminating
atoms at the edges. We also found that all of the self passivating
reconstructions localize electron wavefunctions at the edge. This
localization may be understood in terms of Anderson localization since
the reconstructed edges further break the periodicity of the underlying
lattice. We further analyzed the width dependence of the bandgap in
the reconstructed armchair GNR's, and found that the gap energy is
twice as large as the gap energy of pristine armchair GNR's of the
same width. Finally, we have shown that the band gap in AC56 terminated
GNR's is more robust against the multiplicity, which is of advantage
for fabrication which inevitably introduces width fluctuations. Taken
together, these results elucidate the nature of quantum transport
in graphene based devices with realistic edges and are thus important
for the development of future graphene based nanoelectronic devices. 

\bibliographystyle{achemso}
\bibliography{Edge_Recon_Refs}

\end{document}